%% file: main.tex
\documentclass[sigconf]{acmart}
\usepackage{pifont}
\usepackage{multirow}
\usepackage{CJKutf8}
\usepackage{rotating}
\usepackage{pdflscape}
\usepackage{tabularx}
\usepackage{amsmath}
\usepackage{array}
\usepackage{booktabs}
\usepackage{caption}
\usepackage{subcaption}
\usepackage{acmart-taps}
\AtBeginDocument{%
  }

\copyrightyear{2025}
\acmYear{2025}
\setcopyright{cc}
\setcctype{by}
\acmConference[CHI '25]{CHI Conference on Human Factors in Computing Systems}{April 26-May 1, 2025}{Yokohama, Japan}
\acmBooktitle{CHI Conference on Human Factors in Computing Systems (CHI '25), April 26-May 1, 2025, Yokohama, Japan}\acmDOI{10.1145/3706598.3713341}
\acmISBN{979-8-4007-1394-1/25/04}

\newcommand{\si}[1]{\out{{\small\textcolor{olive}{\bf [*** Si: #1]}}}}

\newcommand{\rev}[1]{{{#1}}}

\usepackage{graphicx}

\begin{document}
\begin{CJK*}{UTF8}{gbsn}

\title[Chinese College Applications - \textit{GaoKao}]{From Scores to Careers: Understanding AI's Role in Supporting Collaborative Family Decision-Making in Chinese College Applications}

\author{Si Chen}
\authornote{Both authors contributed equally to this research.}
\affiliation{
  \institution{Information Sciences, University of Illinois Urbana-Champaign}
  \city{Champaign}
  \state{Illinois}
  \country{USA}
}
\email{sic3@illinois.edu}

\author{Jingyi Xie}
\authornotemark[1]
\affiliation{%
  \institution{College of Information Sciences and Technology, Pennsylvania State University}
  \city{University Park}
  \state{Pennsylvania}
  \country{USA}
}
\email{jzx5099@psu.edu}

\author{Ge Wang}
\affiliation{%
  \institution{Institute for Human-Centered Artificial Intelligence (HAI), Stanford University}
  \city{Stanford}
  \state{California}
  \country{USA}
}
\email{gew@stanford.edu}

\author{Haizhou Wang}
\affiliation{%
  \institution{College of Information Sciences and Technology, Pennsylvania State University}
  \city{University Park}
  \state{Pennsylvania}
  \country{USA}
}
\email{hjw5074@psu.edu}

\author{Haocong Cheng}
\affiliation{
  \institution{Information Sciences, University of Illinois Urbana-Champaign}
  \city{Champaign}
  \state{Illinois}
  \country{USA}
}
\email{haocong2@illinois.edu}

\author{Yun Huang}
\affiliation{
  \institution{Information Sciences, University of Illinois Urbana-Champaign}
  \city{Champaign}
  \state{Illinois}
  \country{USA}
}
\email{yunhuang@illinois.edu}

\renewcommand{\shortauthors}{Chen et al.}

\begin{abstract} This study investigates how 18-year-old students, parents, and experts in China utilize artificial intelligence (AI) tools to support decision-making in college applications during college entrance exam- a highly competitive, score-driven, annual national exam. Through 32 interviews, we examine the use of Quark GaoKao, an AI tool that generates college application lists and acceptance probabilities based on exam scores, historical data, preferred locations, etc. Our findings show that AI tools are predominantly used by parents with limited involvement from students, and often focus on immediate exam results, failing to address long-term career goals. We also identify challenges such as misleading AI recommendations, and irresponsible use of AI by third-party consultant agencies.  Finally, we offer design insights to better support multi-stakeholders' decision-making in families, especially in the Chinese context, and discuss how emerging AI tools create barriers for families with fewer resources.
\end{abstract}



\maketitle

\input{1-intro}

\input{2-relatedworks}
\input{3-background}
\input{4-method}

\input{5-RQ1}

\input{6-RQ2}

\input{7-discussion}
\input{8-conclusion}


\bibliographystyle{ACM-Reference-Format}
\bibliography{sample-base}

\input{9-appendix}

\end{CJK*}
\end{document}

%% file: 1-intro.tex
\qquad\quad 万般皆下品，唯有读书高。---- (北宋) 汪洙

\aptLtoX[graphic=no,type=html]{}{\begin{quote}}
\aptLtoX[graphic=no,type=html]{\qquad\quad}{}All pursuits are inferior; only studying is esteemed-- Wang Zhu (around the Year 1111)

\aptLtoX[graphic=no,type=html]{\qquad\quad}{}\textit{(Translation of Traditional Chinese Slang)}

\aptLtoX[graphic=no,type=html]{}{\end{quote}}

%
%
%
%

\section{Introduction}
On June 7, 2024, 13 million 18-year-old Chinese students sat for the GaoKao (高考) exam across the country, setting a new record since the exam's reinstatement in 1977~\cite{globaltimes2024gaokao}. The GaoKao is one of China's most crucial annual national exams, serving as the primary entrance test for colleges. Of these 13 million examinees this single year, two-thirds will not secure spots in undergraduate programs. Institutions predict the number of GaoKao examinees will continue to rise until around 2035~\cite{pires2019gaokao}. Henan Province in China leads with 1.36 million students taking the exam in 2024 alone~\cite{globaltimes2024gaokao}. It is a highly selective and score-driven exam, taken by almost all high school graduates and others with equivalent academic qualifications. GaoKao is vital for students, shaping their future education and career paths. It has also been considered a key mechanism for promoting social mobility and fairness, with stories of empowering underprivileged students rising through its ranks~\cite{pires2019gaokao}. The exam reflects the society belief that ``knowledge changes fate,'' continuing China's ancient emphasis on education~\cite{harrell1987concept}, as seen in the millennia-old saying in the manuscript opening ``Only studying is esteemed,'' dating back to a time when Columbus had yet to discover the New World.


Unlike many colleges in the US, GaoKao admissions are predominantly based on exam scores from the closed-book tests taken between \rev{June} 7-9. Students submit their college applications after receiving their scores, with a limit on the number of schools and programs they can apply to (decided by the regional government). The admission process is fully automated through a centralized national system, with no human intervention required or essays involved. In recent years, to help students select more satisfactory schools and programs, the number of maximum college applications allowed by the regional government has gradually increased—from 6 options in the past to as many as 90+ in recent years~\cite{globaltimes2024gaokao}. However, this increase in options has also led to a heavier workload for students when filling out their applications. Given that most students are around 18 years old and the college application process is both complex and high-stakes, the decision-making has increasingly become a collaborative effort involving the entire family~\cite{tsegay2016students}.

\begin{figure*}[!h]
    \centering
    \includegraphics[width=0.80\linewidth]{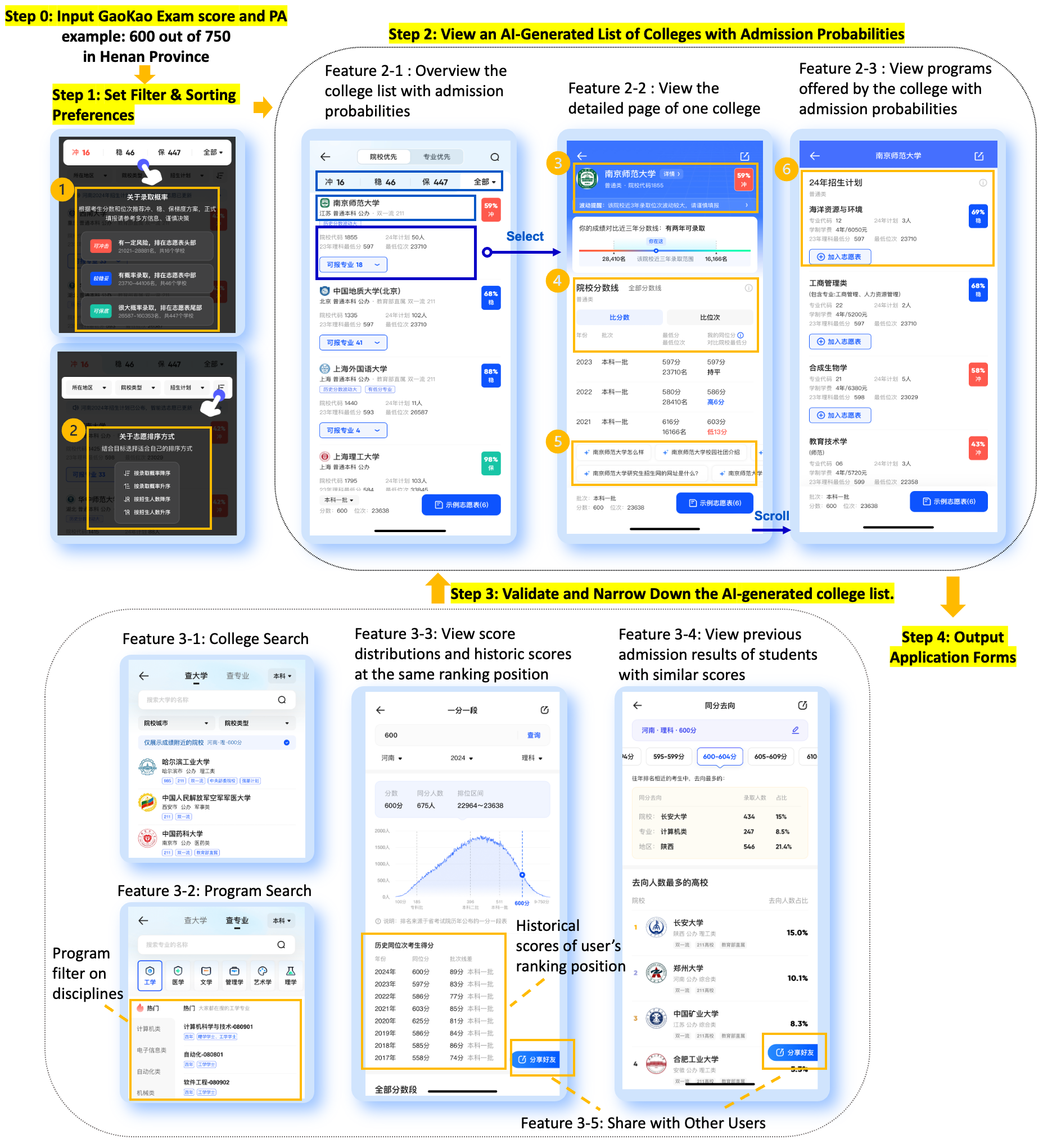}
    \caption{\rev{An example of using Quark GaoKao for a user from Henan Province with a GaoKao score of 600, and ranking position of No. 23638. 
    After the user inputs their scores and PA (Step 0), they can set filter and sorting preferences (Step 1) to view an AI-generated list of colleges with admission probabilities (Step 2). Users may refine the AI-generated college list through features that helps validating and narrowing down (Step 3). Then, users may output the application form to be submitted to the government system.}
    Highlighted sections: \ding{172} Filter options for colleges based on the probability of admission: 16 reach schools are hard to get into (in red), 46 target schools where users' chances are good (in blue), and 447 safety schools that are easier to get into (in green). \ding{173} Four sorting options for colleges: probability of admission descending/ascending, planned admission number descending/ascending. \ding{174} Name, code, category, and overall probability of admission to the selected colleges. \ding{175} Cutoff scores and cutoff ranking positions in the past years for the selected colleges. \ding{176} AI-generated FAQs regarding the selected colleges. \ding{177} Admission plan of a program in the selected college, with details including the program name, planned number of admissions, program code, tuition per 4 years, cutoff exam score and cutoff ranking position last year, and the probability of admission to the program. (Note: the data used in this image is only for demonstration purposes. \rev{Screenshots were taken in August 2024.})}
    \Description[Illustration of Quark GaoKao application]{
    This figure contains 9 screenshots illustrating the 4 major steps when using Quark GaoKao app. To get started (Step 0), users input GaoKao exam score and PA. The example shown is based on 600 out of 750 scores in Henan Province. Step 1 is set filter and sorting preferences. Two screenshots illustrate the filter and sorting options, respectively. The first screenshot has number 1 labeled, and the second screenshot has number 2 labeled. Step 2 is to view an AI-generated list of colleges with admission probabilities. Three screenshots illustrate feature 2-1, 2-2, and 2-3, respectively. Feature 2-1 shows a list of colleges, where each item has the information of the corresponding college on the left and the probability of admission on the right. Clicking on a college transits to Feature 2-2. The next screenshot for Feature 2-2 illustrate the detail page of a school, having the information of the school at top (labeled as 3), followed by the illustration of the cutoff score, followed by the list of cutoff scores in the previous years (labeled as 4), followed by the AI generated questions (labeled as 5).
    Scrolling down this page transits to feature 2-3 on the next screenshot, which is the list of program offered by the school (labeled as 6).
    Step 3 is validate and narrow down the AI-generated college list, illustrated by four screenshots representing five features from 3-1 to 3-5. Feature 3-1 and 3-2 are the search page for schools and programs, respectively. Feature 3-3 is the score and ranking distribution page, having the distribution diagram in the top and historical data in the bottom. Feature 3-4 is the page of admission result of previous student with similar scores, having the top 3 schools in the top, and a list of schools with detailed data in the bottom. Feature 3-5 is share with other users, which is a button on the bottom right of the last two screenshots. Step 4 is output application forms.
    }
    \label{fig:quark_ui}
\end{figure*}

With recent advancements in technology, AI tools are now available to help students fill out their college applications based on their scores. These tools are widely available and offer similar functionalities. This manuscript focuses on how parents, students, and experts (N=32 in total) view and use a widely utilized free AI tool called Quark GaoKao (夸克高考) via interviews. \rev{Tools like Quark GaoKao are branded as AI tools and provide data-driven suggestions that come with probability (e.g., 80\% chance this student can be admitted to X program in college A). Among the many similar AI tools available, we selected the Quark App for this study because it is free to use and widely used in China. Public reports highlight its prominence, with the app leading the industry in user growth and surpassing 100 million uses of its AI features during the GaoKao season. Introduced five years ago, Quark GaoKao gained popularity in the past two years after launching comprehensive GaoKao services, including live streaming before exams, and post-exam college applications. Quark GaoKao is a feature of the Quark Browser, that uses AI search for efficient information retrieval and ensures accuracy through significant manual data correction.} The main function of Quark GaoKao is to generate a list of recommended schools and their respective admission probabilities based on students' GaoKao scores from the \rev{June} 7-9 exams, historical data, and their preferred regions and cities, etc. The sample screenshot is Fig \ref{fig:quark_ui}- \rev{Step 2}. As far as we know, very limited HCI research has been conducted in this unique scenario, where AI-based decisions have long-lasting impacts,  are about students/teens, and the majority of Chinese students grow up experiencing this scenario.  
In this research, we answer the following research question: 
\begin{itemize}
    \item \textbf{RQ1}: How do different stakeholders (students/children, parents, experts) use AI \rev{ tools that recommend schools and programs} to support decision-making in higher education admission? 
    \item \textbf{RQ2}: What are the challenges and opportunities in using \rev{technology} to support decision-making in higher education admission?
\end{itemize}

We make the following contributions:
\rev{1) new empirical understanding of AI tool use in the context of college applications; and 2) novel implications to family-centered design of AI tools for collaborative decision-making.  
First, we examined the use of AI tools, such as Quark GaoKao, that predict the likelihood of student admissions to college programs. Our focus was on understanding how stakeholders interpret and act upon these predictive results in their decision-making process, rather than analyzing the technical methodology behind the likelihood calculations. These tools are predominantly used by parents during the college application process, however, with limited engagement from students. While the tools focus heavily on scores, they fail to address students' long-term career goals. Our study identifies challenges and design opportunities, including misleading AI features, irresponsible use by consultant agencies, and the need to prioritize children's voices in shaping their long-term career development. Second, this work contributes timely implications to Family-centered design of AI \cite{workshop}, an emerging area of research in HCI that emphasizes  the development of technological solutions tailored to the diverse needs and dynamics of families. We also identified synergies--such as communication and relationships, students' well-being, and self-actualization--and tensions, including information asymmetry, emotional asymmetry, and generational differences between students and parents when using  AI tools for collaborative decision-making. Additionally, we  highlighted external and internal factors, such as technology literacy, internet access, screen time, and socio-technical resources that need to be considered to address these tensions and achieve synergies between students and parents (two stakeholders) in family-centered design. 

Our findings are analyzed from sociotechnical lenses, revealing significant implications. While AI is often designed to enhance educational access, our study uncovers how it can inadvertently deepen inequities. Factors such as parental awareness, access to resources, technological proficiency, and access to necessary tools intersect to influence how families engage with AI, often reinforcing existing social disparities.
}






%% file: 2-relatedworks.tex
\section{Related Work}

\subsection{Technology for Career Development}

In recent years, technology has played an increasingly pivotal role in supporting individuals in making informed career choices, evolving from basic digital assessments to sophisticated AI-driven systems~\cite{venable2010using}. Early developments in this domain centered on digitizing traditional paper-based career assessments, such as the Myers-Briggs Type Indicator (MBTI)~\cite{pittenger1993measuring} and Holland's Occupational Themes (RIASEC)~\cite{de1997five}, transforming them into online platforms that offer more nuanced, data-driven insights. These platforms assess personality traits, skills, and preferences, automating the process of matching users with potential career paths~\cite{yanti2020development,rao2020use}. As the field has advanced, AI has further revolutionized career decision-making technologies. AI-driven systems, such as Pymetrics~\cite{wilson2021building}, analyze user behavior through neuroscience-based assessments and employ machine learning to predict suitable career paths based on cognitive games and psychometric data~\cite{tonny2022ml}. Similarly, LinkedIn's Career Explorer~\cite{case2013linkedin} leverages large datasets on job roles, skills, and career trajectories to provide personalized insights on skill gaps and career transitions, enabling users to make data-informed decisions about whether to remain in or pivot from their current fields.

In the context of supporting students in making career decisions, platforms like Naviance~\cite{deslonde2018technology} and College Board's BigFuture~\cite{arevalo2023pathways} offer guidance by analyzing students' academic performance, interests, and career aspirations to provide personalized college and career recommendations. Similarly, Kuder Navigator~\cite{landherr2019relationship} helps students explore career options through assessments and interactive tools that align career aspirations with relevant educational pathways. Additionally, AI-driven chatbots, such as CareerVillage~\cite{kurniawan2024sustainable} and AI4All~\cite{judd2020activities}, allow students to ask career-related questions, receiving advice from professionals and data-driven insights about potential career paths. These tools enable students to explore career options early on and make informed choices regarding college majors, internships, and future careers, tailoring their education to market demands and personal aspirations. 

As for college admissions, previous research on automated systems in holistic college admissions has examined how algorithms support the evaluation of various applicant attributes, such as academic performance, extracurricular activities, and personal essays~\cite{talkad2018making}. These studies emphasize the potential of automation to process large volumes of applications efficiently while striving for fairness and thoroughness. However, concerns about algorithmic bias, transparency, and the inability to replicate human judgment remain~\cite{alvero2020ai,kreiter2013proposal}, and much of this research has focused on Western contexts, leaving other regions and admission processes underexplored. In China, the only basis for college admissions is the GaoKao score, making the process highly quantitative~\cite{tsegay2016students}. As a result, local AI tools in China are designed to provide updated statistics and accurate admission probability calculations. In contrast, college applications in the US and many other countries are more versatile, with factors such as grades, standardized test scores, extracurricular activities, and personal essays all contributing to admission decisions. Tools like Clab AI~\cite{clabappai2024} and Kollegio~\cite{kollegio2024} cater to this holistic process by focusing on content generation. For example, Kollegio helps users determine which activities align with the colleges and majors they are applying to and suggests personal statement topics that fit the users' backgrounds. Additionally, the tool assists users in writing activity descriptions and essays using AI. This contrast highlights the different roles AI plays in supporting college admissions across regions, reflecting the varied decision-making criteria used in each context.

In this paper, we focus on the score-driven national college entrance exam, GaoKao, to examine how technology can assist students in making informed career decisions under high-stakes conditions. Unlike other educational systems that may allow for a more holistic assessment of students' capabilities, GaoKao's singular focus on exam scores creates a highly competitive and pressure-filled environment where students' futures hinge on their performance. Given this context, the integration of technology becomes especially significant. By addressing this exam-driven context, we highlight the critical role technology plays in supporting students' academic and career outcomes.

\subsection{Family Joint Use of AI Applications}

Recent research on family decision-making highlights how AI applications are becoming integral to collaborative family choices, from mundane tasks to more complex decisions~\cite{yu2024exploring,guo2019review}. Families often use AI-driven systems such as recommendation algorithms and virtual assistants to facilitate joint decisions on everyday activities, like selecting entertainment, managing household chores, or organizing family schedules~\cite{tamura2013recommendation,xu2018family}. These systems provide tailored suggestions based on individual preferences, which family members negotiate to arrive at collective decisions. However, the use of AI in these contexts also introduces new dynamics into the family decision-making process.

Power dynamics within families play a significant role in how AI is used and interpreted. Recent studies have shown that while AI systems offer neutral, data-driven insights, they do not eliminate the social hierarchies that influence family decisions~\cite{su2014family}. Parents, for example, often exercise greater control over how AI recommendations are considered, especially in decisions that impact the family unit as a whole~\cite{beneteau2020parenting}. In this context, AI tools can reinforce existing power structures, with parents filtering or overriding recommendations to maintain authority over final decisions, particularly in areas like finance or education~\cite{stone2024dawn}. However, AI can also shift these dynamics. Research suggests that children's familiarity and comfort with AI technologies can sometimes challenge traditional power hierarchies, giving younger family members greater influence in decision-making, particularly in technology-driven tasks~\cite{shifflet2016adolescent,wang2023data}. For instance, children often act as intermediaries, assisting parents in navigating AI systems, which can redistribute decision-making power in more tech-savvy families~\cite{wang2023data}. Family-oriented educational tools that use AI to recommend learning paths can foster more democratic engagement by providing personalized suggestions for each family member to consider in discussions~\cite{bajwa2024parenting}. These technologies offer shared points of reference that facilitate more equal participation, though how this unfolds often depends on underlying family dynamics~\cite{shifflet2016adolescent}. \rev{These work aligns with the growing body of research on family-centered design~\cite{10.1145/3613905.3636290}, which emphasizes the importance of technologies that support the negotiation of diverse values, needs, and preferences among family members while fostering cooperation in their use~\cite{cagiltay2023child}. Such designs highlight the critical role of understanding the social dynamics and situated experiences of families to create effective technological solutions for the digital age. By addressing challenges such as conflicting goals, facilitating communication, and promoting shared engagement, family-centered interaction design offers valuable insights for fostering long-term connection and collaboration within families~\cite{dalvand2014family}.}

In this work, we focused on this important scenario of family joint engagement with AI applications to explore the potential dynamics and shared decision-making when it comes to important academic or career-related decisions. Specifically, we examined how AI tools, such as recommendation systems for educational or career planning, could influence the way families navigate these significant choices together.


%% file: 3-background.tex
\section{Context: AI Features for Supporting GaoKao Application}
\subsection{Background of College Application in Mainland China}
The National College Entrance Examination in mainland China, commonly referred to as the ``GaoKao,'' is one of the most significant national exams in China. It serves as the admission examination for universities and colleges across mainland China and is taken by high school graduates or other candidates with equivalent academic qualifications.
\rev{
For the vast majority high school graduates who are willing to pursue undergraduate level education, GaoKao is the only viable option to apply for college, and therefore it is extremely competitive.
} 
\\

\begin{figure*}[th]
    \centering
    \includegraphics[width=0.9\linewidth]{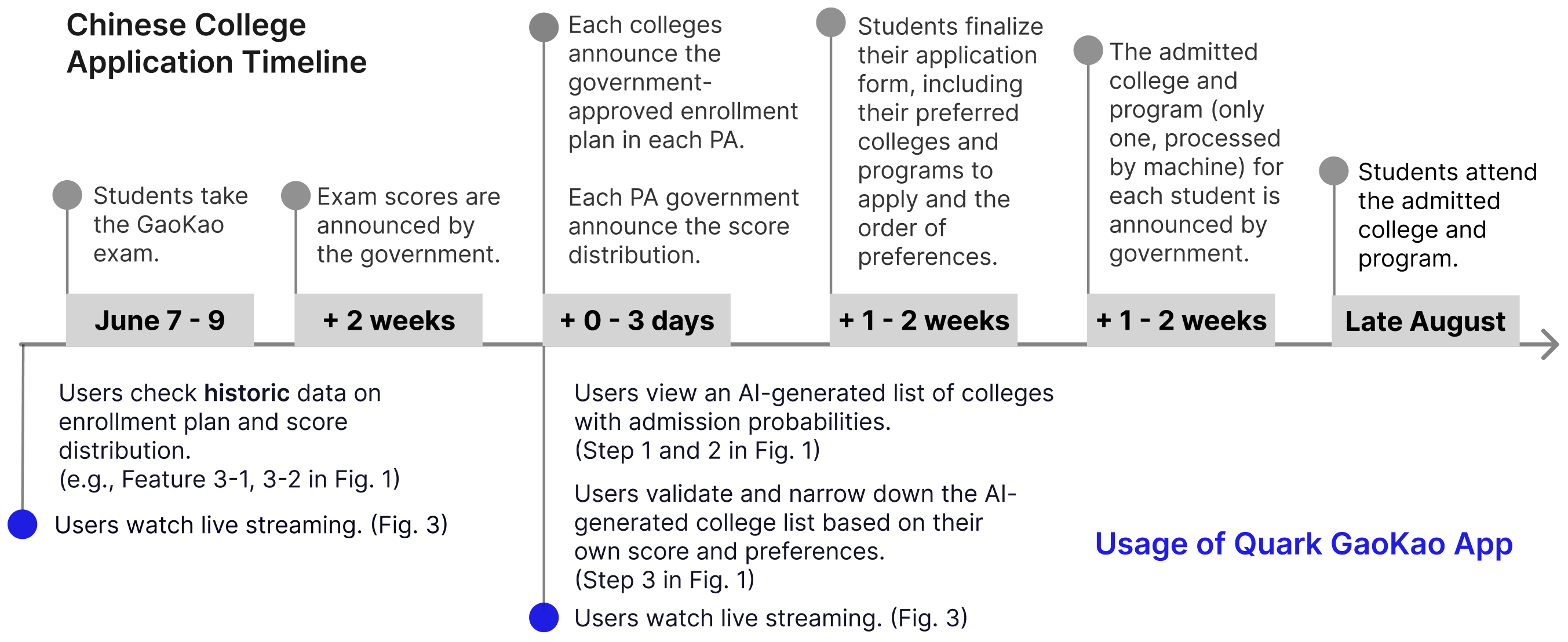}
    \caption{\rev{The college application process and timeline in mainland China, as well as the usage of Quark GaoKao app throughout the process.}}
    \Description{This figure is an illustration of Chinese college application timeline and usage of Quark GaoKao app on the timeline. There are six squares in the middle representing six key dates of the application process. The first is June 7 to 9, when Students take the GaoKao exam. The second is two weeks later, when the exam scores are announced by the government. The third is 0 to 3 days later, when each college announce the government-approved enrollment plan in each PA, and each PA government announce the score distribution. The fourth is another 1 to 2 weeks later, when students finalize their application form, including their preferred colleges and programs to apply and the order of preferences. The fifth is another 1 to 2 weeks later, when the admitted college and program (only one, processed by machine) for each student is announced by the government. The last is late August, when the students attend the admitted college and program. For the usage of Quark GaoKao app, the first support starts before the first date on the timeline, when users check historic data on enrollment plan and score distribution, and users watch live streaming. The second support is at the third date on the timeline after enrollment plans and score distributions are announced, where users view an AI-generated list of colleges with admission probabilities (Step 1 and 2 in Figure 1), user validate and narrow down the AI-generated college list based on their own score and preferences (Step 3 in Figure 1), and user watch live streaming (Figure 3).   
    }
    \label{fig:gaokao_process}
\end{figure*}

\rev{Although GaoKao is a national event, students are taking the exam locally in the PAs (PA refers to province-level administrative divisions, which include a province, autonomous region, municipality, or special administrative region. In 2024, there are 31 PAs that administer GaoKao) they live in.}
Since 2004, the Chinese government has allowed PAs to prepare their own GaoKao exam questions, leading to a system where exam activities and scores \textbf{vary by PA}, and scores are not equivalent and transferable across PAs. 
\rev{Besides,}
the number of students taking GaoKao exams also varies between different PAs,
\rev{due to the unevenly distributed population.}
For example, while Henan province, one of the PAs where many of our participants came from, has approximately 1.36 million students participating in GaoKao in 2024, the municipality of Shanghai only has approximately 58,000 students for the same year.
\rev{
The independent administration of the exam and the disparity in number of students consequently make each college across the country to set different target enrollments and admission cutoff scores for students from different PAs, based on the intra-PA exam scores and rankings.
}

\rev{The college application in mainland China is processed by the Chinese government and is based on the student's GaoKao score, ranking position in their PA, and their application form. Unlike college application process in some other countries, a student can only be admitted to one program (major) in one college assigned by the government. 
Therefore, completing college applications in mainland China is a non-trivial task, which is essentially a strategic decision-making process that can be described by various Game Theory models~\cite{chen2014chinese}. Students have roughly one week after the score is released to complete the application form and submit to the government. The application form is essentially \textbf{an ordered list of colleges and programs} based on the student's preferences. 
The submitted forms are processed by machines designed by the government to assign students to the college and program based on their GaoKao scores and their application form, with students who had higher scores being processed first and assigned to the hightest ordered program and college that have not reached planned admission number. Different students' applications are processed in order from the highest to the lowest score. 
A summary of the college application process in mainland China is described in Fig. \ref{fig:gaokao_process}. For detailed explanations of each step, please refer to Appendix \ref{sec:gaokao_steps}. }

Since 2014, the reformed GaoKao (following Government Policy \footnote{\url{https://www.gov.cn/zhengce/content/2014-09/04/content_9065.htm}}) has gradually expanded to more PAs, bringing significant changes to the college application process. One key change is the significant \textbf{increase in the number of allowed college applications} per student, with some PAs now permitting up to 112 choices, compared to the previous limit of 6-10 choices. These regional differences reflect how the reform is tailored to local policies and needs, with some PAs offering more flexibility than others.

\subsection{Feature Summary of Quark GaoKao App to Support College Application}








In response to the various challenges associated with college application decisions based on the scores from GaoKao exam, a growing number of technological tools have been developed to assist families in making more informed choices. Among them, Quark GaoKao (夸克高考) (referred to as Quark in the remainder of the paper) is the one of the most popular apps used by Chinese high school students to assist their college application.
While it has been launched for over six years~\footnote{\url{https://www.quark.cn/article005}}, it gained more attention from the students and their families in recent year after its integration of AI technology.
Due to its free-of-charge nature, Quark GaoKao has been widely used by students and their families, and all participants in this study have used it as one of their tools to support the college application process. The primary goal of Quark GaoKao is to assist in completing the college application process after GaoKao scores are released, which involves selecting and ranking preferred colleges and programs. 
\rev{
The core usage of Quark can be summarized in four main steps to support students and their families in finalizing their application form within one week after receiving their scores, with different features supporting. Fig. \ref{fig:quark_ui} illustrates a typical usage of Quark GaoKao. To get started, Quark requires the users to input PA and GaoKao scores.

\noindent ■ \textbf{Step 1: Set filter and sorting preferences.} In this step, users set filter and sorting preferences, such as selecting a preferred region of the colleges, the type of colleges, and number of applicants to be accepted by a college.

\noindent ■ \textbf{Step 2: View an AI-generated list of colleges with admission probabilities.} In Step 2, users view an AI-generated list of colleges with admission probabilities calculated based on their scores and relevant historical data. Users have the following features available during Step 2:

\begin{itemize}

    \item \textit{Feature 2-1: Overview the college list with admission probabilities.} Quark shows the AI-generated list of colleges and programs with admission probabilities on an overview page. Quark also arranges the colleges into three categories based on different probabilities: reach (red), target (blue), and safety (green). Users can swipe left or right to prioritize either colleges or programs (by inputting their preferred majors visually similar to Feature 3-2). All of our participants reported using the default ``prioritize colleges'' option.
    
    \item \textit{Feature 2-2: View the detailed page of one college.} Users may select a college from the overview page to review the details of the selected college, including the college name and code (to be filled on the application form) (\ding{174} in Fig. \ref{fig:quark_ui}). Users may also review the historical admission data, including the cutoff score and ranking (\ding{175} in Fig. \ref{fig:quark_ui}). In addition, there are AI-generated FAQs (\ding{176} in Fig. \ref{fig:quark_ui}), which are answered by AI upon clicking. None of our participants mentioned reading AI-generated FAQs, as they were perceived as less relevant to increasing the likelihood of score-driven admission. 

    \item \textit{Feature 2-3: View programs offered by the college with admission probabilities.} By scrolling down the detail page, users can see the programs offered by that college, along with the probabilities of admission into each program.



\end{itemize}

\noindent ■ \textbf{Step 3: Validate and narrow down the AI-generated college list.} In this step, users may validate and narrow down the AI-generated college list to be submitted in the application form. Quark provides several features to support this step:

\begin{itemize}
    \item \textit{Feature 3-1: College search.} Users may search for any college to learn more about their programs.

    \item \textit{Feature 3-2: Program search.} Similarly, users may search for a program they are interested in, which can be filtered by disciplines. They are ranked in end-user search popularity.

    \item \textit{Feature 3-3: View score distributions and historic scores at the same ranking position.} Users may review their ranking position in their PA on the score distribution of the current year. Based on their ranking position, Quark will also display the scores that received the same ranking in previous years in that PA. Since the exam questions are different each year, showing the scores for the same ranking position could better help users understand the historical data.

    \item \textit{Feature 3-4: View previous admission results of students with similar scores.} Based on the user's scores and ranking positions, Quark will display the popular colleges which previous students in similar ranking positions were admitted. This list is sorted by the percentage of students going to a college, with the most popular colleges listed first.

    \item \textit{Feature 3-5: Share with other users.} Users may share the information in Feature 3-3 and 3-4 with other users, such as their family members and advisors.

\end{itemize}

\noindent ■ \textbf{Step 4: Output application form.} After 
Quark can then output the pre-filled application form with the selected colleges, which can be submitted to the government system.

Additionally, Quark offers the following features that can be accessed before GaoKao score is released and throughout the application process without impacting the AI-generated college lists or calculated probabilities. 

\begin{itemize}
    \item \textbf{Historic data on admission statistics.}
Quark provides features that could help students gather various historical admission statistics of a college, including cutoff scores and enrollment plans. Users may search for such information through college search and program search (Feature 3-1 and 3-2 in Fig. \ref{fig:quark_ui}). Such data is particularly valuable to users seeking to practice filling out application forms by referencing their personal historical data.

\item \textbf{Live streaming.}
One commonly used feature by our participants is live streaming (as shown in Fig. \ref{fig:livestream}, which is available daily and hosted by experts in the college application process, covering different topics with various streamers in each session. For example, the live streams may discuss suggestions to fill application forms or programs with better employment rates. Users can ask questions to the streamers during the live streams. 

\item \textbf{MBTI personality tests.}
Other features offered by Quark include MBTI personality tests to help students understand themselves, though none of our participants mentioned using this feature.

\end{itemize}

\begin{figure}[th]
    \centering
    \includegraphics[width=0.99\linewidth]{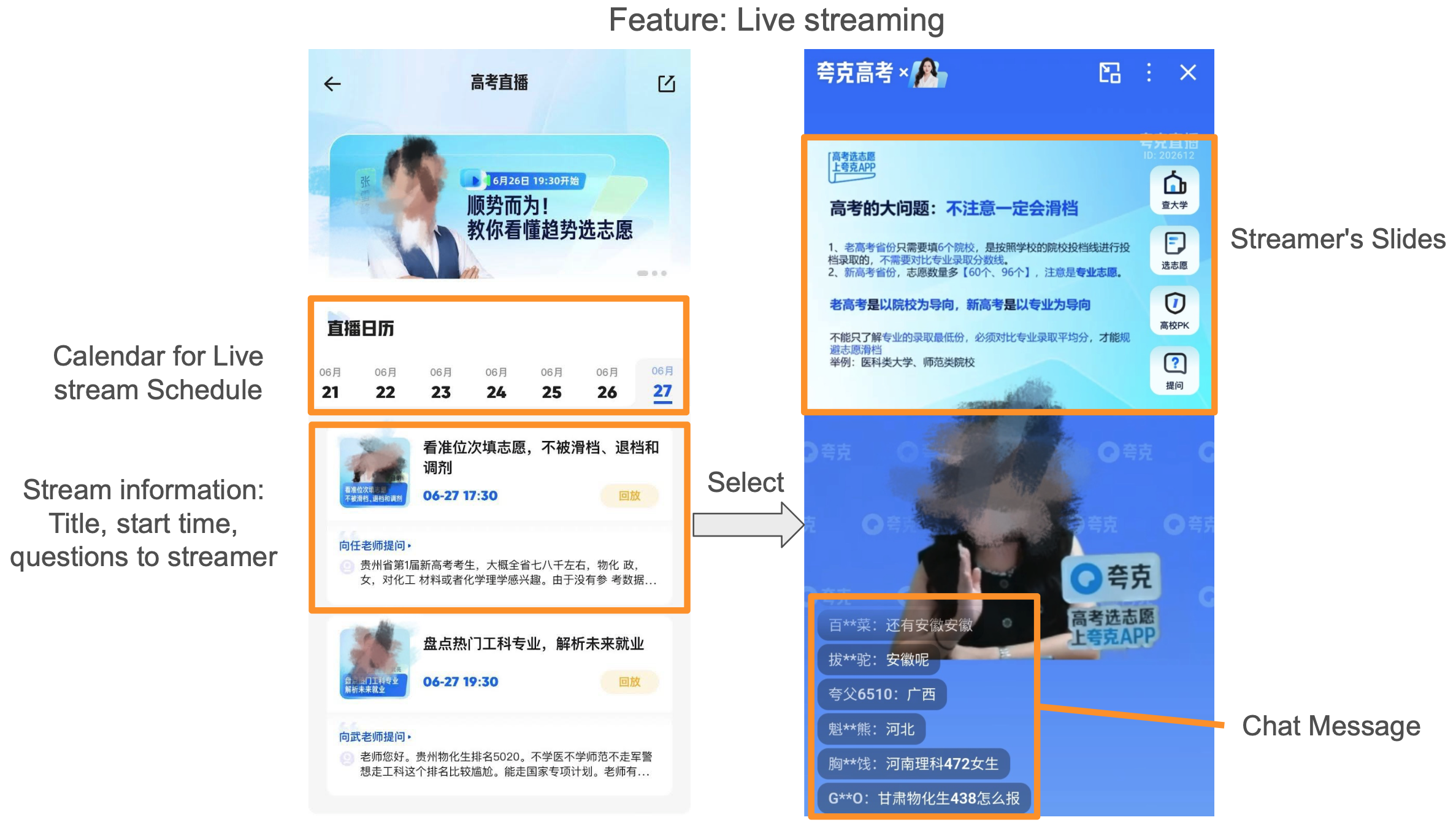}
    \caption{\rev{The live streaming feature offered by Quark is widely used by our participants, particularly by parents seeking knowledge before the GaoKao exam. Users may check the upcoming live-streaming schedule and select the topics they are interested in. They may also rewatch a past livestream. Users may select a live stream from the schedule to join. In this example, the stream was about strategies to order the list of colleges and programs in the application form to avoid the application being rejected or skipped (if all colleges on the form were full when the application was being processed, the form would be skipped or rejected, leading to the students having no colleges to attend). Users may interact with the streamer through chat messages to ask questions during the live stream. In this example, viewers were asking for suggestions in chat by providing their PA and GaoKao scores.}}
    \Description{There are two screenshots showing the live streaming feature offered by Quark GaoKao app. On the left is a screen showing the schedule of live streams. Users can choose a date on a calendar to view live streaming schedule on that date in the middle of the screen. Below that, they can review the stream information of each session, including title, start time, and sample questions to the streamer. Users can click on a session to join the stream. The screenshot on the right is taken during a live stream. Streamer's slides are shown on the top. The streamer is in the middle. The chat messages are shown on bottom right. 
    
    }
    \label{fig:livestream}
\end{figure}
}

%% file: 4-method.tex
\section{Method}

We interviewed 32 participants, including paired family-children, individual family members, individual children, and experts, to gain different stakeholders' perspectives of AI usage in score-driven Chinese college applications.
This study is IRB-approved. \rev{All student participants were 18 years or older at the time of the study and reflected on their experiences using the tool during their high school years. We obtained oral consent approved by a U.S. institution, in alignment with the authors' current affiliation. The participants represented the first cohort to utilize the tool during the GaoKao exam and university application process. }

\subsection{Participants}
We recruited a total of 32 participants (20 males and 12 females) via word-of-mouth and posts on social media platforms. The family groups comprised 7 pairs of family members (e.g., father, mother, and uncle) along with their children, 6 individual children whose families had limited engagement in the college application process, and 4 individual family members whose children had limited involvement in the college application process. 
We made efforts to maximize regional diversity during recruitment.
Additionally, we recruited 8 experts who specialized in guiding or assisting families with college applications. These experts comprised middle school teachers, full-time and part-time counselors, and university admission office staff.

All the family members and children have utilized at least one AI tool for college applications. For family groups, 23 out of 24 participants or their children attended GaoKao and completed the college applications between 2023 and 2024,  during the rapid growth and popularization of AI tools in GaoKao application process. One exception is K3, who attended GaoKao in 2021 and utilized AI tools to verify her choices in recent years.
All the experts have either used AI tools for college applications or have interpreted results from such tools.
Table~\ref{demo_info} and~\ref{demo_info_expert} present their demographics. 
Each participant received a \$10 gift card for their time and effort, unless they preferred not to be paid.


\subsection{Procedure}
We conducted 32 semi-structured interviews via WeChat (a Chinese messaging and call app), each lasting between 20 and 40 minutes in Chinese. Each interview is facilitated with one or two researchers. \rev{To ensure participation from individuals across various regions of China and to minimize technological requirements, we conducted voice chats instead of video calls. This approach reduced the need for high-speed internet and eliminated the necessity of traveling, making the study more accessible to geographically diverse participants.}

First, we asked participants for basic information. 
For children and family members, we inquired about the region and year they (or their children) attended GaoKao due to the variation in subjects and admission policies across different PAs, as well as the tier of schools to which they (or their children) were admitted. For experts, we asked about the regions where they have assisted families with college applications.

Second, we delved into participants' experience with AI tools used in 
the college application process. We asked about which AI tools they have used, which features were utilized, when they started using the tools, and how long they have been using them. 
Subsequently, we explored the impact of these AI tools on their decision-making process. This included the factors that AI tools considered or failed to consider, participants' trust in these AI tools, their satisfaction with the tools, and suggestions for improvements.

Third, we explored potential collaboration or conflicts between children and family members. This contained the process of collaboration, reasons for potential conflicts, and strategies for fostering collaboration or mitigating conflicts.

Finally, we asked whether participants used other resources besides AI tools for decision-making. 
For those who used additional resources, we gathered information on what these resources were, how they combined the use of AI tools with other resources, and their comparisons between AI tools and other resources.

\subsection{Data Collection \& Analysis}

Audio recordings were made for all interviews with participants' consent, and the interviews were conducted in Chinese. The first two authors transcribed all recordings into text, ensuring that participants were fully de-identified. Since all authors are native Chinese speakers, we were able to seamlessly conduct data analysis using the original transcripts.

We used a bottom-up approach in our qualitative data analysis. We performed inductive thematic analysis~\cite{braun2006using} on the transcribed interviews individually, used open coding to develop initial codes, and generated themes and subthemes through iterative collating and grouping. The themes and subthemes were reviewed and finalized during weekly meetings with all authors.

%% file: 5-RQ1.tex
\section{RQ1-Quark Usage by Different Stakeholders}

In this section, we present our findings related to research question 1, focusing on how various stakeholders -- students/children, parents, and experts -- utilize AI to support decision-making in higher education admissions. To clarify participant references, we use specific codes: PF denotes the parent in a parent-child pair, PK refers to the child in a parent-child pair, F represents individual parents, K indicates individual children, and E stands for experts. For more detailed demographic information on our participants, please refer to Table 1 in the Appendix.

\subsection{Stakeholders' Agreement: App Met Score Optimization Goal Yet Career Development Goals}


Participants leveraged various AI features to optimize their use of GaoKao scores in selecting colleges. They accessed historical data on cut-off scores (Feature 3-3 in Figure~\ref{fig:quark_ui}) and learned about newly introduced majors (Feature 3-2 in Figure~\ref{fig:quark_ui}) to expand their options for consideration (PF1-5, PF7, PK1, PK3, PK6, K1, K2, K4, F13, F14, E2, E3, E5).
Additionally, they (PF4, PF7, PK1, K4-6, F11-14, E7) employed AI tools to guide them in narrowing down a balanced mix of colleges, categorized into ambitious (reach), reasonably attainable (target), and safer (safety) options (Step 1 in Figure~\ref{fig:quark_ui}), shaping a strategic application approach. 
Another significant use of AI involved assessing the probability of acceptance at different colleges (Feature 2-1 in Figure~\ref{fig:quark_ui}), which was important for finalizing their application choices
(PF1, PF2, PF5, PF7, PK1, PK3, PK6, K1-6, F11, F12, F14, E1, E7).

The importance of long-term career prospects significantly influenced the college selection process. Twenty-one participants (PF1-7, PK3, PK6, K1, K2, K6, F11-14, E2-4, E6, E7) marked it as a crucial factor in their final decisions. 
Typically, they defined effective career development primarily through two criteria: first, by choosing majors that aligned with \rev{their (or their children's)} interests at colleges; and second, by securing stable employment like government positions, teaching, policing, or medical roles, or pursuing further academic research after graduation. 
\rev{This pattern reflects the practical considerations unique to the Chinese job market, where certain professions offer particularly stable career trajectories.}
However, participants reported that current AI tools are inadequate in offering essential information or support in these aspects, highlighting a gap in the technology's capability to aid in decision-making related to long-term career planning.


Eight participants (PK1, PK4, PF4, PF5, F13, E3, E5, E7) raised concerns that AI tools fail to incorporate personal preferences and individual interests when suggesting colleges or majors. Instead of considering what the children likes or prefers, these tools predominantly use GaoKao scores to determine potential admissions to specific majors or schools. 
PF4 noted that although AI tools effectively utilize the scores to match children with prestigious colleges, the majors that the children truly desires are often out of reach in the recommended colleges, according to historical admission scores. 

\begin{quote}
    \textit{``The AI system suggests many schools, but most don't meet our needs. It might recommend a school and tell you which majors you can get into, but often these are less popular choices or not something you actually want, it's picked because your score just meets the cutoff. For majors you're really interested in, you're often unlikely to get in based on past scores.''} (PF4)
\end{quote}
\rev{
This illustrates how AI's score-centric approach creates a mismatch between recommendations and student aspirations. The system's rigid reliance on GaoKao cut-off scores means it often suggests majors solely based on score eligibility rather than students' interest, ultimately diminishing the tool's practical value for career planning. This limitation is particularly significant in the Chinese higher education context, where major selection is directly tied to GaoKao performance, and switching majors after admission is uncommon.
}



PK1 and PK4 echoed that this limitation is particularly acute for students in the humanities-track, who face a narrower range of available majors at Chinese colleges. The AI tools tend to recommend majors that are less commonly studied, leading to a mismatch between the recommendations and the students' actual interests. Consequently, students spend more time browsing through options they aren't interested in. 
In a related instance, F12 encountered a situation where AI attempted to personalize recommendations by considering her child's personality test responses. Unfortunately, these recommendations proved inaccurate, as they did not reflect her child's true interests, highlighting a critical flaw in the AI's approach to personalization.



Seven participants (PK3, PK6, K2, K6, E2, E3, E6) mentioned AI's inability to take account for long-term career development when guiding students post-graduation. K6 pointed out that AI recommendations are solely based on present academic performance, neglecting how well the suggested schools or majors might align with the student's future career goals. This limitation suggests that while AI is helpful in the immediate selection process, it lacks the foresight necessary to advise on how these decisions could influence a student's career trajectory in the long run.
\rev{The disconnect between AI recommendations and long-term career planning was articulated by K6.}

\begin{quote}
    \textit{``AI's limitation compared to human guidance is that it doesn't consider your future development. It only focuses on your current situation, calculating which schools you can apply to based on your scores. Beyond that, it offers little else. AI needs to improve by offering more comprehensive services, like future career planning and other long-term considerations.''} (K6)
\end{quote}
\rev{
This observation highlights a gap in current AI systems, where they excel at processing GaoKao scores and historical admission data, but they fail to incorporate career trajectory planning. K6's experience reflects a broader pattern among our participants who found AI tools inadequate for connecting present academic choices with future professional opportunities. 
}

\subsection{Parent-Led Data Triangulation around AI Usage and Trust Building}
\label{data_triangulation}

While AI tools provide valuable insights into the college selection process, skepticism remains among participants (PF1, PK6, K6, F11, F14, E1, E3), particularly regarding the reliability of AI-generated acceptance probabilities. This skepticism is rooted in the belief that these probabilities are derived from historical data, which might not accurately reflect current trends or shifts in the popularity of majors.

To address these concerns, participants (mostly parents) leveraged data from additional sources to validate and supplement on AI's results. These sources include live streaming about college admissions, guidebooks for GaoKao college applications, and consultations with agencies. PF7 articulated the reasons for data triangulation to cross-verify information on AI tools and make informed decisions.

\begin{quote}
    \textit{``In the matter of GaoKao, which is a critical juncture for the child, 80\% or 90\% of parents would not completely trust any single AI tool or consultant agency... (AI) can achieve perfection, but still, parents won't trust it fully; they can't.''} (PF7)
\end{quote}

\paragraph{Live Streaming}
Live streaming features on these AI platforms has emerged as a pivotal tool for disseminating information about GaoKao process and subsequent career pathways, including detailed introductions to various majors and course curricula. Sixteen participants (PF1, PF2, PF4, PF6, PF7, K1, K2, K4, K5, K6, F13, F14, E2, E5, E7, E8) reported that they or their parents engaged in watching live streaming on these AI platforms to gain GaoKao-related information. 
Live streaming transcends geographic and socioeconomic barriers, allowing parents from diverse backgrounds to gain specialized information from experts in an accessible, interactive, and real-time manner. 

\rev{
Through these streams, experts provide detailed explanations about major requirements, curriculum content, and career prospects -- information that isn't readily available through AI's data-driven recommendations. 
}
PF4 highlighted how live streaming can clarify misunderstandings about \rev{majors.}
\begin{quote}
    \textit{``(Live streaming) breaks the information cocoons... Like with GaoKao, there are some aspects that are difficult for us to pay attention to, and we actually need someone to clarify these for us. As ordinary people, we're unclear about suitable majors and career paths. For instance, some humanities majors are named `a' but teach `b'.''} (PF4) 
\end{quote}
\rev{
This expert guidance through livestreaming helps families better understand the nuances of major selection and reduces the risk of students enrolling in majors that don't align with their expectations or career goals. However, the commercial nature of these livestreaming platforms introduces issues with information quality and objectivity. 
Since these platforms operate on a for-profit model (PF7, E8), the content may be influenced by commercial interests, potentially influencing parents' and children's decisions (K2, E6). We will further explain this in Section~\ref{socialmedia}. 
}

\paragraph{Book Reading}
Six participants (PF2, PF7, K1, F12, E3, E5) utilized books, such as the official guidebook for GaoKao College applications (Figure~\ref{fig:cutoffbook}) and supplementary books for Youzhiyuan, to verify and complement the information provided by AI tools.  
Despite their less user-friendly format for quick data retrieval compared to AI tools, books offer in-depth explanations of admission policies that AI data often fails to capture. PF7 noted the value of books in more thoroughly explaining \textit{``the nuanced differences between sequential, parallel, and early admission processes at colleges, along with specific admission rules and scoring variations among colleges.''}
Moreover, books like the official guidebook for GaoKao College applications are appreciated for their authoritative presentation of data. P12 mentioned the reliability of information sourced from books: \textit{``the data on books is definitely more scientific and more realistic, because Quark may not be as comprehensive.''} 

\paragraph{Consultant Agencies}
Six participants (PF2, PF7, K3, K4, E2, E8) indicated that parents consulted with agencies, that provide paid services to support the GaoKao application process. 
This consultation process typically began either before or after GaoKao.
According to E2 and E4, the consultant services typically cost around 5000 RMB (700 USD) per student.
These consultant agencies complements AI tools by incorporating professional insights and discussions into the decision-making process. Initially, agencies employ AI tools to screen potential college options based on students' personalities, academic scores, and regional preferences (PF7, K3, E2, E8). 

While AI tools lay the groundwork by narrowing down options, the value of consultant agencies lies in their subsequent step: personalized discussions with parents and students to finalize college selections through offline, one-on-one support. This process involves discussing each considered major, the future career paths associated with them, and the subjects that would be studied. It ensures that choices align well with the student's interests and capabilities. 
PF7 highlighted the benefit of this approach: \textit{``I seek professional advice from consultant agency to ensure my child understands the major's content, related subjects, and potential career paths in detail.''}
This personal touch allows for significant communication and exploration, where consultants engage directly with students to explore individual needs and give related explanations or suggestions.

\subsection{Experts Explains AI Overlooks in Considering Localisation Policy}

\rev{
While parents and students navigate AI limitations through data triangulation, experts in our study revealed additional concerns specific to the GaoKao admission system. With their extensive experience in guiding students through the college selection process, these experts possess deeper understanding of admission scores and policies compared to parents and students. Three of them (E3, E5, E6) reported encountering significant inaccuracies and gaps in AI tools. Their concerns center on AI tools' handling of regional admission policies, enrollment quota changes, and specialized program requirements. This expert perspective helps explain why stakeholders, particularly parents, seek multiple information sources beyond AI recommendations (Section~\ref{data_triangulation}) to make informed decisions about college and major selection. 
}

E3 and E5 noted discrepancies in the cut-off scores listed on Baidu (a Chinese search engine platform) compared to those in the official guidebook. Such errors have led to their reduced trust in certain AI applications. E5, drawing on his expertise in Tianjin, observed: \textit{``There are some deviations in cut-off scores, the number of students enrolled, and whether the majors actually enrolled students last year, which is somewhat biased from the data on official guidebook.''}

Experts (E3, E5, E6) also indicated that AI tools often lack information on specialized programs. E6 highlighted the overlooked advantages of niche colleges like Shijiazhuang Railway Institute, which may offer better career prospects in contrast to more generalized colleges -- a nuance typically absent in AI tools. 
Moreover, E5 critiqued AI tools for their lack of timely updates regarding changes in admission policies, such as adjustments student enrollment numbers or the discontinuation of certain majors.



\subsection{Student's Limited Involvement with Whole Workflow -- Crossing Out Options}


\rev{
While our previous findings showed how parents actively engage with AI tools and various information sources, a contrasting pattern emerged regarding children's participation. 
}
Nine participants (PF1-4, PK6, F11, F14, E1, E7) pointed out that children had limited involvement in the college selection process. Children often lacked a clear understanding of their own interests, leading them to primarily eliminate options from a list initially compiled by parents or experts rather than actively selecting preferences.

Participants articulated that children's involvement tends to be more about rejection of dislikes rather than an enthusiastic pursuit of interests. This was attributed to a limited exploration of potential career paths, which left many children uncertain about their interests and future directions. For example, PF3 shared insights into the child's decision-making process. 

\begin{quote}
    \textit{``He just says what he doesn't want to do. As for what he wants to do, he's not very clear. He mentions a general direction towards engineering, but whether it's automation, computer science, robotics, or something else, he can't specify because he hasn't had exposure to such work. So, he can only specify a general direction of what he wants to do, but what he doesn't want to do is clearer. He doesn't want to study psychology, doesn't want to deal with machines, doesn't want to deal with people.''} (PF3)
\end{quote}
\rev{
This illustrates how the lack of practical exposure to different fields creates a decision-making pattern where children find it easier to identify what they don't want rather than what they do want to pursue. 
}




\rev{
The challenge of children's limited involvement appears to be exacerbated by current AI tools' limitations. E5, drawing from extensive counseling experience, highlighted how AI systems fail to provide the guided exploration that children need. 
}

\begin{quote}
    \textit{``The ideal AI should meet the same logical requirements as a GaoKao counselor. For instance, if a student comes in knowing nothing, you must be able to guide them step by step to uncover their questions... Thus, I believe that the biggest challenge we currently face with AI systems is how to help a student who has no clear ideas to develop their own thoughts and discover their needs.''} (E5)
\end{quote}
\rev{
This points out that while AI systems can process academic data and generate recommendations, they lack the capability to guide students through the exploration of potential career paths that align with their interests, as well as to adequately understand and respond to children's needs. This limitation explains why student's involvement remains primarily reactive rather than proactive in the college selection process. 
}

%% file: 6-RQ2.tex
\section{RQ2-Technology Challenges and Opportunities}

In this section, we present our findings related to research question 2, which explores the challenges and opportunities associated with using AI to support decision-making in higher education admissions.


\subsection{Children Limited by Internet Access and Interaction with Parents} Participants reported that high school students had limited opportunities to communicate with their parents due to attending boarding schools or managing heavy coursework prioritized by both teachers and parents. In most PAs, especially outside Beijing and Tianjin, students were restricted from using phones or the internet during school hours, and some families enforced these rules even after school. This lack of access made it difficult for children to explore career options or gather information beyond their school curriculum. In one positive case, a student attending a boarding school with limited internet access found alternative ways to develop their interests by reading books about various fields and careers. This early exploration helped ease the college application process significantly, as the student felt well-prepared and confident about their choices, and his parents were not significantly involved in the decision-making process. As shown below. 

\begin{quote}
    \textit{``When I was in school, through reading some books and magazines, I found that cybersecurity within the field of computer science really attracted me. I had decided to study computer science quite early on compared to others around me, but the specific school was only determined after the exam results came out. The general direction was already set before applying, allowing me to complete the application without any issues.''} (K2)
\end{quote}


Other children faced challenges when their parents tried to involve them in the application decision-making process only after the GaoKao exams. Exhausted from the intense academic pressure leading up to GaoKao, these children felt disconnected from the decisions they were expected to make, and how to process various data sources in decision making resulting in a lack of engagement and motivation. Correspondingly, many parents felt uncertain about how to effectively involve their children, as they were unclear about their children's preferences, strengths, and communication preferences. 

To address communication issues with her son which she expected, PF7 hired a consultant agency. She specifically selected a consultant with persona (e.g., an outgoing male in his 20s from the same region) whom she believed would better communicate with her son. She then conveyed her ideas to the consultant, asking them to relay her thoughts to her son. However, with the application deadline approaching (usually 3-7 days between score announcement and application submission), parents often took the lead in ``\textit{getting it done}.'' Echoing PF7's experience, E8 who used to work at consultant agencies found that most parents who come to agencies are seeking emotional support due to the short decision time available due to ``\textit{High Pressure and Tight Deadlines}'' (Chinese Slang- 时间紧任务重). Limited access to technology in boarding schools and coursework load made it hard for students to explore career options or participate in decision-making. Many parents took the lead, hiring consultants to meet deadlines.

In most families in our study, explicit conflict between parents and children was rare. While parents acknowledged some misalignment in priorities (e.g., region, major), they felt they managed these differences well, even though their children didn't express any conflicts experience to the researcher. Only one child mentioned her frustration with her mother's reliance on consultant agencies, feeling it reflected a lack of trust in her (K3). Experts' observations supported both parents' and children's views. When children lead decision-making, they feel more responsible and satisfied with the outcomes. In contrast, when parents dominate, children are more likely to shift blame when the children face obstacles. These findings suggest centering children's voices could foster responsibility and trust, though this dynamic is not yet fully realized. The technology could potentially mediate this process.

\subsection{Tensions Between Parents and Students in Decision-Making}

\rev{
In addition to the communication issues, we identified tensions between parents and students stemming from divergent expectations about decision-making timing, as well as conflicting priorities regarding institutional prestige versus major selection. 

The immediate post-GaoKao period emerged as a particularly stressful time, highlighting the disconnect between parents' urgency and students' readiness to engage in the decision-making process. K3's experience illustrates this tension. 

\begin{quote}
    \textit{``The obvious conflict started right after I finished GaoKao. I just wanted to rest at first, but my parents kept pushing me to quickly start looking at college preferences. This made me really resent the whole process of filling out preferences. I felt it wouldn't be too late to wait until the scores were released.''} (K3)
\end{quote}

K3's experience reveals a tension between parents and students over the temporal demands of college application. While parents pressed for immediate engagement with this process, students like K3 needed a period of rest after the intense GaoKao. This misalignment in preferred timing led to a tension, with K3 developing resentment toward the college application task itself. These temporal pressures are further complicated by parents' own educational backgrounds and limited familiarity with higher education systems, as E2 explains:

\begin{quote}
    \textit{``The majority of Chinese parents have not received higher education, so they feel a sense of insecurity about schools. Because of this, they rely on AI tools and external assistance.''} (E2)
\end{quote}

E2's observation discloses how parents' limited experience with higher education leads them to rely on external validation sources, including AI tools and consultant agencies. While parents rely on external sources to compensate for their knowledge gaps, our findings reveal a deeper generational divide in understanding future career development. Students, having grown up in a digital era, are more attuned to how AI and computing technologies are reshaping industries and career opportunities. In contrast, parents often focus on traditional criteria of success like institutional prestige. This mismatch in understanding future career demands creates another parent-child conflict.

\begin{quote}
    \textit{``Parents might not have a good understanding of academic majors. They know the reputation of schools like Tsinghua and Peking University are prestigious, right? But once you get in, it's not certain what you'll actually end up studying. Meanwhile, the child, who is exposed to more current information, might know they want to study something like computer science and understand its earning potential after graduation. This creates a conflict: should they aim for a top-tier school with an uncertain major, or settle for a slightly less prestigious school but secure a strong major?''} (E5)
\end{quote}

E5's insight demonstrates how parents' focus on institutional prestige can overlook the practical implications of major selection on the students' future careers. This disconnect, compounded by AI tool's emphasis on score optimization over career development, can inadvertently perpetuate decision-making tensions between parents and students.

}

\subsection{Triangulating Data Requires Extensive Time, Literacy and Social Capital} 

Conversations between parents and children in our study started as early as the 10th grade when students made decisions about their course selections, which would ultimately limit their major options. For example, students who didn't take physics were excluded from certain engineering fields like mechanical engineering. Families such as PF1 (with a PhD degree) and PK1 engaged in in-depth discussions throughout 10th to 12th grade, carefully considering their choices over time. The most intensive period of preparation occurred in the last three months of 12th grade, with families working daily to narrow down a long list of around 200 schools recommended by AI tools to a more focused shortlist of about 20 schools. While AI provided the initial recommendations, much of the narrowing down required significant human effort. PF1 anticipated that the final score would fluctuate, and he input different anticipated scores (e.g., 580, 600, 620) into the tool to prepare various plans. PK1 explained how PF1 used a paid tool, similar to Quark, to provide additional data and resources that helped triangulate the AI-generated list and involve her idea in the process. This process of cross-referencing data ensured that their decisions were well-informed and less stressful as the final deadline approached.

\begin{quote}
    \textit{``Starting in March or April, my father (PF1) estimated my score around 600 and mapped out options for various scenarios: 620-630, 600, or even 550-580. He researched each school's environment and admissions policies across these score ranges. After the exam results were out, we discussed my mother's wish for me to stay in Beijing and my preference to avoid language majors. With each conversation, we narrowed down choices, and by the time we decided on my current program, many unsuitable options were already eliminated.''} (PK1)
\end{quote}



Experts and parents echoed that narrowing down school options is a process largely driven by human decision-making, with the predicted likelihood of acceptance often proving less useful than expected, as reflected in RQ1 findings. This process is extensive and demanding, requiring significant effort to compare school details, assess personal preferences, and navigate complex data during Step 3 in Fig. \ref{fig:quark_ui}. For students with mid-range scores, the challenge is even greater due to the sheer number of applicants with similar scores. For example, as shown in Figure \ref{fig:quark_ui} Feature 3-3, around 800 students scored exactly 600 out of 750, and in the most crowded range -- 450 out of 700 -- there were over 1,800 students with identical scores. This high level of competition makes it difficult for students to stand out, adding complexity to the already challenging process of selecting schools. E8 remarked that various family-specific factors further complicated the process of narrowing down school options, such as literacy, power dynamics, and AI tools were unable to account for these differences effectively. 

\begin{quote}
    ``\textit{When it comes to serving parents, there are many factors involved. There is anxiety about making applications under temporal pressure, and there are many other issues such as literacy diversity, and communication style. From my perspective, as someone working in education technology, I find it quite disheartening because it's not effective. We originally thought this product could solve at least 70–80\% of these issues, but in reality, it solves less than 10\%.}'' (E8)
\end{quote}


The parent noted that live streaming is viewed as more understandable by families with challenges to use tool AI features to narrow down school lists, who often lack direction and the ability to process complex data and information, such as numbers and policy details. In contrast, parents with more field knowledge find the information less valuable, as explained in detail in Section \ref{socialmedia}. PF2, a parent with a middle school education level, relied heavily on live streaming, taking notes for two years while working over 10 hours daily.

\begin{quote}
    ``\textit{It's hard for me because, with my lower education level, I struggle to understand written info, like what a mechanical engineering major involves or its career paths. I live in a small town with small agencies that help with college applications. They consult families like mine, who don't understand these things well. While I try to help my child and understand his preferences, agencies and teachers mostly focus on scores and ask a few questions, but they don't know our kids as well as we do.''}  
\end{quote}


\begin{quote}
    \textit{``
    It (livestreaming) was somewhat helpful (in increasing education access). For example, many don't know about the supplemental admissions process (a policy), but now, with info online, like Douyin (TikTok) streams, more people are aware. We also received a big admissions plan book, and I went through it with my child, comparing all the data. I was worried about picking the wrong major or entering the wrong code. We also used Quark to compare last year's scores for our decision.}'' (PF2)
\end{quote}



Even with live streams that seem to provide easier-to-understand information than AI features, experts and parents expressed concerns that families with more social capital are better equipped to navigate the challenges around data works. These families are more likely to find ways or secure external technical and human resources or invest time themselves to manage the process. Social capital is the network of relationships between people that allows a society to function effectively. As demonstrated in the quotes below:

\begin{quote}
    ``\textit{The more complex the rules (for AI and GaoKao and AI for GaoKao) are, the more advantageous they are for people with resources and wealth. Wealthy and powerful individuals, as well as those with resources, have the means to study and navigate these complexities. On the other hand, the simpler the rules, the more they benefit the poor.}'' (E2)
\end{quote}

\begin{quote}
    ``\textit{The complexity of the rules raises two issues: whether families are willing to spend money on school preference selection and whether they guide their child early, like in the first or second year of high school. Many laborer parents are away from home and can't offer guidance, and they may lack the experience to advise on choosing science or subjects like chemistry or biology. In the past, decisions were often arbitrary, but now, when picking a subject like chemistry, you need to consider how well your school teaches it compared to the provincial standard.}'' (E2)
\end{quote}

\subsection{Mixed Influence of Technology in Steering Families Toward Popular Majors}\label{socialmedia} Experts argue that application features often prioritize major popularity, which tends to steer families toward certain majors, such as computer science. For example, in Quark, many families in our study relied on Features 3-1 and 3-2 in Fig. \ref{fig:quark_ui}. In these features, academic programs are ranked by popularity, indicated by a small fire icon (e.g., engineering, medical school, humanities; vertically within the engineering field: computer science, automation, software engineering). Much like the advice given on social media, these app features encourage users to focus on what is trending rather than exploring more personalized or interest-based options. As a result, families often make decisions based on popularity rather than a deeper understanding of the fields their children might enjoy or excel in, missing the opportunity for more personalized decision-making. 

In addition to popularity-driven tool features, both social media platforms and applications like Quark also have live streaming sessions where many live streamers, offering free advice to families through live connection calls, often claim expertise but was observed to provide uniform information to most viewers who call. While there are more experienced streamers, most lack the qualifications needed to offer tailored guidance, frequently repeating the same general advice observed by experts and parents.  PF1, who holds a PhD, observed the streamer fitting the same suggestion to different children using the same template and questioned the qualifications of streamers on live streaming platforms. Streamers also prompt families to sell access codes to paid application support tools similar to Quark and consultation agencies to make money, as explained in Section \ref{agency}. 

\begin{quote}
    ``\textit{The primary role of live streaming is to guide and attract viewers. If you follow them for a while, you realize they often repeat the same one or two points. To access more in-depth information, you have to purchase their paid services.}'' (PF1)
\end{quote}


Both experts and parents found streamers frequently repeat what most families choose, without providing personalized options. A parent critically reflected that streamers following popular decisions could lead to a more competitive environment, where many children end up pursuing the same field. The parent noted that live streaming is viewed as more ``beneficial'' for families with fewer resources, as it provides some direction when they have none, helping parents feel reassured that have not made the wrong decision though not the best decision for their children.

\begin{quote}
    ``\textit{The more ***(streamer's name) says something is popular, the more I feel like what he said can't be relied on. If everyone rushes to a field, the competition will be the toughest... For different social classes, like people who don't have jobs or farmers, going to school is about securing employment. Listening to some of these(livestreaming) ideas might be helpful, but you still have to work hard to have a career. For us in the big companies, I feel many things ***(streamer's name) say is misleading. I am from the chemistry field, I think this field is great because [various reasons from personal experience and observation]... my colleagues and supervisors are all in this industry, if you want to make it easier for your child to find a job, you definitely need to plan ahead for them to work in your own field.}'' (F14)
\end{quote}



\subsection{Misleading AI Claims and Deferred Accountability of College Consultation Agencies}\label{agency} Six families in our study hired consultation agencies to assist with decision-making alongside AI tools, using them to triangulate data. Most consultation sessions are one-on-one family for a couple of hours after the score is announced. At the time of the study, families were satisfied with their college admissions outcomes and generally content with the services provided by the agencies. However, experts pointed out a troubling trend: many agencies falsely advertise their use of novel algorithms and unique data sources to generate superior school application lists -- claims that are misleading, as the relevant data is publicly available through government websites.

Experts further emphasized that the true impact of college applications is long-term and cannot be fully assessed at the time of college acceptance. Despite this, agency contracts often limit accountability, leaving families without a way to hold them responsible for future outcomes. This deferred accountability enables agencies to make misleading AI claims and offer less effective recommendations without facing immediate consequences. In essence, AI is used as a deceptive hook by agencies to attract parents, with no clear mechanism to challenge or disprove these misleading claims or suggestions. Nearly all experts noted that the admission policy is shifting towards allowing learners to submit more applications, reducing the likelihood that they will be left without an offer from any school. As a result, consultants are less likely to encounter strong dissatisfaction from the families in their work. 

\begin{quote}
    \textit{``Given the current admission policies, it's almost impossible for a family to be completely dissatisfied. Just think about it: 96 applications are allowed in Shandong PA, 96 in Hebei PA, and 112 in Liaoning PA -- it's all about ranking, right? Consultants help families pick 96 out of more than 20,000 options, rank them according to preferences, and that's it. The one (school) you get into may not be the one most satisfied with, but at least not something you dislike at that moment. It's easy money to make.''} (E8)
\end{quote}

Five experts and parents challenge the claim that most tools, including the Quark GaoKao application, are truly AI-driven. Instead, they argue these are merely big data tools that do not generate new information or recommendations. The experts emphasize that these tools still heavily depend on the individual consultant's ability to interpret the data and offer personalized guidance to families, particularly addressing the temporal pressures of the tight college application timelines. 

\begin{quote}
    \textit{``Agencies themselves aren't capable of providing consultants with significant benefits... It (agencies and apps) have turned into pure marketing. They manufacture anxiety, and buying their product (service and novel AI tools) makes you reliant on these so-called teachers, doesn't it? I want to use technology to solve problems, but right now, they're just focusing on marketing, like live streaming (to hook families to buy their products), which isn't sustainable.''} (E8)
\end{quote}

%% file: 7-discussion.tex
\section{Discussion}
\rev{Technology has a complex role in family collaboration during the college application process—a high-stakes, time-sensitive decision with delayed outcomes. Educational platforms such as Quark emphasize popular majors, limiting meaningful career exploration, especially for resource-limited families. Although Quark offers features intended to support understanding of student interests, such as MBTI and ``prioritize major'' in Feature 2-1, very few participants were aware of these features or understood their relevance to collaborative decision-making.  Below, we discuss insights for designing education technology in family contexts and unfold social barriers that hinder tools from fulfilling their promise of improving educational access.

}

\begin{figure*}[th]
    \centering
    \includegraphics[width=0.82\linewidth]{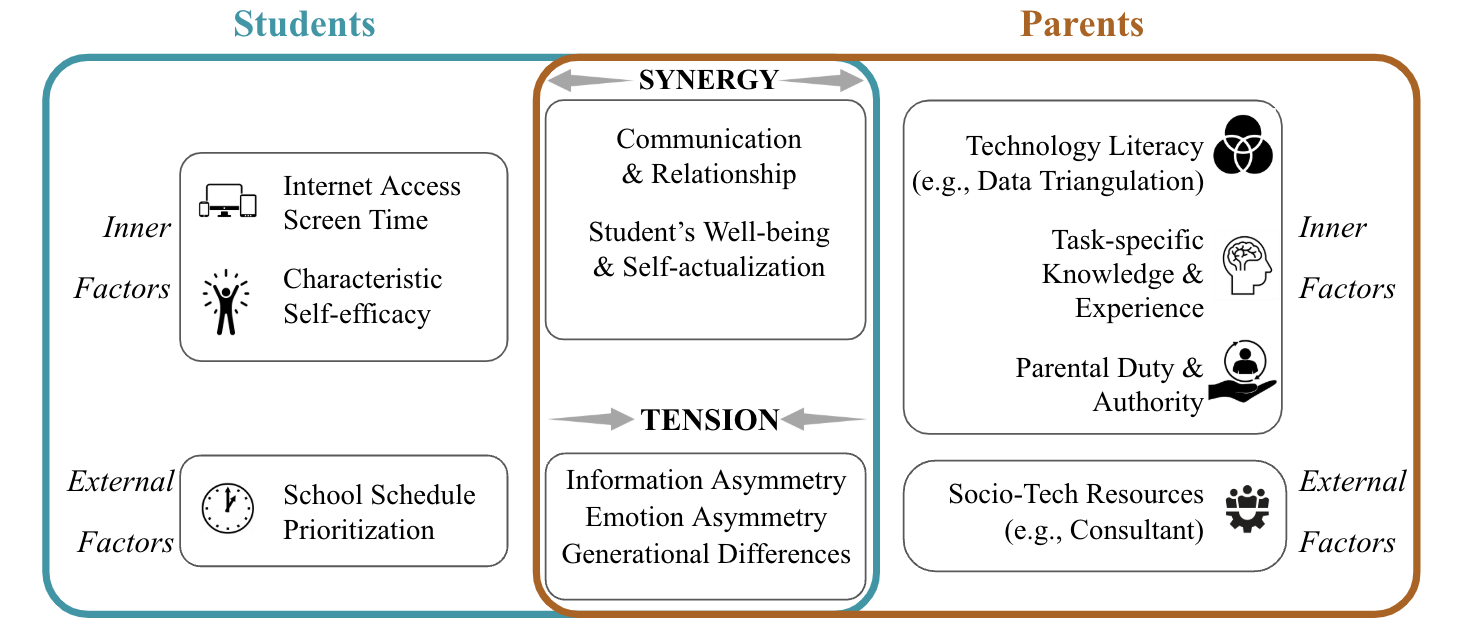}
    \caption{\rev{Summary of Findings on Parent-Student Dynamics. Internal and external factors influence the design of technology that aims to reduce tension and foster synergy between parents and students. }}
    \Description{}
    \label{fig:sum}
\end{figure*}

\rev{
\subsection{Implications for Designing Education Technology in Family Context}

Technology is increasingly pervasive in family life, creating opportunities to design solutions that align with the diverse needs and dynamics within families. This calls for the line of work around family-centered design (FCD) that extends user-centered design principles by focusing on the interconnected and interdependent nature of family members, who share living environments and maintain long-term relationships. Unlike individual user scenarios, FCD emphasizes multi-user participation and inclusive processes, ensuring tools address the collective experiences, emotional transactions, and shared goals of family units~\cite{workshop}. 

Our study builds on FCD by empirically examining how families navigate the GaoKao process, uncovering both opportunities for synergy and sources of tension. Synergies emerge through communication and relationships, where parents and student collaboratively align their goals and learn from one another. This dynamic not only supports students' self-actualization but also helps families balance short-term academic goals with long-term career aspirations. Various factors that impact synergy and tension from our findings are summarized in Fig. \ref{fig:sum}. We consider our table a starting point for exploring family dynamics, and it is not exhaustive to capture all relevant factors. Notably, tensions often stem from informational and emotional asymmetries, as well as generational differences -- challenges that future technology could address more effectively:

\textit{Informational Asymmetries}: Similar to prior research on family decision-making dynamics~\cite{lackman1993family}, these were also evident in our findings. Parents often had greater access to resources, leveraging professional knowledge, data literacy, and AI tools to validate decisions via guidebooks and live streams. In contrast, students, constrained by limited internet access and school demands, relied heavily on parents to interpret and manage information. This gap raises critical opportunities for future education technology design that fosters equitable access to information and supports collaborative decision-making among parents and students. For instance, systems could draw on previous designs to encourage shared responsibility by involving students in collecting and analyzing information~\cite{10.1145/3613904.3642344,10.1145/3544548.3581522}, even with limited time and internet access. Features such as offline access to curated resources, concise data summaries, and interactive tools can empower students to interpret data independently. Additionally, guided prompts or activities could facilitate collaborative evaluation of information, fostering mutual learning and engagement.

\textit{Emotional Asymmetries}: Prior research has shown that parents often take a protective role in high-stakes decision-making, prioritizing student's emotional well-being while absorbing the stress and complexity of such processes~\cite{workman2015parental,allen2014parental}. Our findings align with this, as parents bore the emotional weight of high-stakes decisions, often turning to consultants or external tools to shield students from the associated pressures. While this approach aims to safeguard students, such a protective approach inadvertently distances students from critical discussions, leaving them less prepared for future decision-making~\cite{10.1145/3476084}. This underscores opportunities for education technology to balance emotional support with preparation for independent decision-making. For example, systems could provide scaffolding tools that gently introduce students to the decision-making process, using simulations~\cite{klein2018role} and interactive scenarios~\cite{barfield2010mind} to help them understand the stakes and implications without overwhelming them emotionally. 

\textit{Generational Differences}: In our findings, parents prioritized stable, prestigious careers like teaching or government roles, reflecting traditional societal norms. In contrast, students, more attuned to technological shifts, showed a preference for fields like AI and computer science. These differing priorities often created tension, especially when parents placed greater emphasis on institutional prestige over the long-term career potential of emerging fields. Such tensions reveal the need for educational technology that facilitates intergenerational dialogue and a shared understanding of career pathways. For example, systems could provide tailored visualizations of labor market trends, integrating data on both traditional and emerging career trajectories, similar to prior designs that use career mapping tools to enhance decision-making transparency~\cite{miller2015career}. Interactive features could allow families to explore and compare the stability, growth potential, and societal impact of various careers, fostering informed discussions.







}

\rev{
\subsection{Unpacking Social Barriers to Technology's Promise in Increasing Education Access}
Our findings reveal that tools like Quark GaoKao have the potential to enhance educational access but are limited by social barriers that influence their use and impact. Specifically, in currently parent-led collaboration, we discuss how parental awareness, availability, abilities, and resources intersect to shape the way families navigate AI tools.

\textit{Varied Parental Awareness of Technology-Driven Values}: Our study reveals that system interactions reflect broader societal norms by  centralizing exam scores and popular majors, steering families toward competitive fields. Awareness of these system values varies across users: some parents and experts recognize the score-driven guidance and its alignment with societal priorities, questioning its tendency to overlook personal interests and long-term goals. In contrast, some resource-limited families often perceive AI recommendations as neutral and objective. Without critically engaging with the underlying sentiments, these families rely on AI outputs and defer to prevailing societal opinions. While these features simplify decision-making, they inadvertently discourage the exploration of diverse educational and career pathways, reinforcing societal norms that equate academic performance with success. Live streaming offered accessible and easy-to-understand knowledge about college applications. However, it was criticized for being misleading, as it often directed students toward popular majors without considering long-term planning for their futures and encouraged impulsive consumption. To address these challenges, future system designs may foster critical engagement~\cite{spector2019inquiry} with AI recommendations (e.g., critical thinking prompts such as, ``How does this option align with your long-term goals?''), encouraging families to question underlying assumptions and explore a wider range of possibilities. Meanwhile, incorporating features that promote long-term opportunities~\cite{ahrari2024ai} and individualized guidance beyond core-based metrics, can help align decision-making with users' evolving goals and contexts. For instance, systems could prompt families to explore alternative pathways with interactive ``What If'' scenarios or career trajectories tied to non-conventional majors. 

\textit{Varied Parental Ability and Availability to Involve Student's Voice}:
Parental resources significantly influence family engagement with AI recommendations. While the system includes features like MBTI with the potential to center students' voices, our participants were largely unaware of their relevance to collaborative decision-making. Some parents relied on paid consultant services to mediate communication with their children, but these services were not affordable for everyone. Some parents have the intention and availability to communicate with their children daily, while others face challenges due to their children attending boarding schools or parents working in other cities away from home. This indicates a critical gap: existing features that aim to center students' voices lack sufficient communication, accessibility, and contextual support to ensure meaningful and equitable use across diverse family contexts. To address this disparity, technology must foster awareness and enhance parents' ability to involve student's voices in a meaningful and sustained way. For instance, synchronous and asynchronous collaboration tools like chatbots or shared planning features~\cite{nihei2024chatbots} could allow students to share preferences independently, ensuring flexible communication despite geographic or time-based constraints.

\textit{Varied Data Literacy among Parents}: Parents' ability to validate AI recommendations using resources like guidebooks, live streams, or consultants is crucial to decision-making. Experts noted that some families can differentiate between original information and interpreted data, such as live streams, which may carry biases. Families with higher data literacy critically evaluate AI recommendations, aligning them with student-parent shared goals and leveraging social capital, while those lacking these skills and resources face inequities. Bridging this gap requires AI tools that can interactively support data interpretation and make essential information accessible to parents and students across diverse literacy levels and resource availability.
}

\rev{\textit{The Impact of the Changing GaoKao Policy on Families} GaoKao remains a highly score-driven system, emphasizing academic performance in shaping students' futures. As mentioned in the introduction, in recent years, the number of maximum college applications allowed by regional governments has increased significantly—aiming to help students select more satisfactory colleges and programs. This is part of the larger policy shift introducing the 新高考模式 (``new GaoKao model''), which has added complexity to the system through frameworks such as the ``3+3'' model.
More specifically, the first batch of reforms was implemented in Shanghai and Zhejiang, followed by the second batch in Beijing, Hainan, Shandong, and Tianjin, which also adopted the ``3+3'' model. Under this framework, students can freely choose three elective subjects from physics, chemistry, biology, history, geography, politics, and technology. The combination of courses they select directly impacts the subjects of the exams they will take, the colleges they can apply to, and the scores required. 

A few participants in our study, from regions under the new model, highlighted that it has significantly increased the complexity of the data-driven system. They noted that students are now required to make critical decisions about their career paths and course selections much earlier in life, often in contexts where limited historical data is available to guide them. These changes also demand earlier involvement from families to avoid exacerbating unforeseen inequities, as children without sufficient support may face greater challenges navigating this system. The evolving GaoKao framework underscores the growing need for technology to focus on career development tools that can guide students and families through these changes, helping to alleviate the pressures of early career decisions and reduce systemic disparities.}

\rev{
\subsection{Limitations} Our study did not include younger students under the age of 18, most of whom have not yet taken the GaoKao exam. Since the GaoKao is typically taken around the age of 18 and marks the culmination of high school education in China, this exclusion represents a limitation and an area for future research. 
Additionally, our sample was regionally concentrated in northern China, which may restrict the applicability of our findings to other areas. Future studies should aim to include participants from southern regions to capture a more diverse range of experiences. Variations in regional policies across China also likely influence the adoption and effectiveness of AI tools, and future research should investigate how these differences shape user behavior and access to such technologies. Finally, the reliance on voice chats, while reducing technological barriers, may have excluded individuals with limited phone or internet access. We conducted interviews and gathered UI screenshots before September 2024, so our analysis excludes Quark versions released afterward. As Quark undergoes frequent updates, future research could explore newer features to provide updated insights. The final limitation is that, although Quark provided AI-generated FAQs (\ding{176} in Fig. \ref{fig:quark_ui}) to facilitate faster consumption of online data and resources, none of our participants reported using it or even noticing the feature. This highlights a need for future research to explore it.}

%% file: 8-conclusion.tex
\section{Conclusions}\rev{This study examines the role of AI tools, such as Quark GaoKao, in the highly score-driven Chinese college application process, via studying multi-stakeholders. While these tools effectively generate application lists based on scores and preferences, they often fail to consider students' long-term career aspirations and personal interests. The parent-led usage of the tool in decision-making highlights gaps in amplifying student agency and fostering longitudinal engagement with AI for career planning. The lack of accountability and transparency in how consultant agencies use these tools further underscores the need for ethical oversight. Our findings contribute to family-centered design by identifying synergies (e.g., improved communication and relationships) and tensions (e.g., information and emotional asymmetry) in parent-student dynamics. Socio-technical factors, such as disparities in technology literacy and access to resources, create barriers for resource-limited families (e.g., access and ability to validate AI recommendations via different data sources). 
}

\begin{acks} 
We appreciate Dr. Sooyeon Lee for her invaluable feedback on the study design, which greatly strengthened the methodological rigor of this work. 
We also thank the anonymous reviewers for their insightful comments and suggestions. 
\end{acks} 

%% file: 9-appendix.tex
\clearpage
\appendix
\onecolumn
\section{Appendix}
\label{appendix}

\rev{
\subsection{Description of GaoKao process} \label{sec:gaokao_steps}
}

The regular admission round in the college application process is the standard process in mainlan China where the majority of colleges and programs admit students after GaoKao. It covers most colleges at various levels. There is also an early admission round that refers typically to military academies, police colleges, teacher training colleges, and special programs like arts and sports. International students have other application tracks. The early admission round and other application tracks are not the focus of this paper.

In the following, we describe the key steps of the regular admission round in the college application process in mainland China in chronological order, as described in Fig. \ref{fig:gaokao_process}.



\noindent ■ \textit{Students take the GaoKao exam.} The national GaoKao exam has been held annually between June 7th to 9th.

\noindent ■ \textit{Exam scores are announced by the government.} Exam scores are usually announced between June 23 and 25 in most PAs. The maximum score is 750 points in most PAs, including both the core subjects and elective subjects from the academic proficiency exams. Each of the core subjects—Chinese, Mathematics, and a foreign language—has a maximum score of 150 points. The three elective subjects chosen by the student are graded and then converted into a score that counts toward the total, with each subject having a maximum score of 100 points. \textbf{Only the GaoKao score is required for the application and serves as the sole basis for admission decisions.} 

\noindent ■ \textit{Each PA government announce the score distribution.} About 0 to 3 days after the exam scores are announced, each PA government announces the score distribution, segmented into 1-point intervals. Students could see their rank within the PA. All historical data can also be found on the government website. It can also be found on Quark. 


\noindent ■ \textit{Colleges announce the government-approved enrollment plan in each PA.} College enrollment plans for each PA are released 2 to 3 days after the announcement of GaoKao exam scores. These plans are developed by colleges qualified to enroll students through GaoKao and are based on the nationally approved annual enrollment quotas and outline the distribution of enrollment slots by program for each PA (a college may have different quotas in each PA). While the enrollment plan of a college for a specific PA may undergo minor adjustments each year, the historic enrollment plans from previous years can serve as a valuable reference for students and can be found on the government website. These plans guide students in selecting their preferred colleges.


\noindent ■ \textit{Students finalize their application form.} \rev{\textbf{Key Supports by Quark.}} In most PAs, students fill out their college applications only after knowing their GaoKao scores. The deadline is usually around one week after GaoKao exam scores are released, althought it might differ among PAs. For example, In Henan Province, where most of our participants are from, the application process in 2024 is conducted in three phases: between June 26 and June 28; between June 30 and July 2; between July 4 to July 6. During each phase, students can apply up to 12 colleges, ordered by their preferences. For each college, they can choose up to 5 programs, ordered by their preferences, and indicate whether they agree to be placed in a different program if their chosen program is unavailable. 


\noindent ■ \textit{The application result (processed by machine) is announced by government.} 
The applications are process solely by machine in each PA in the regular admission track following the process found in the government website: a priority by score ranking, the equal status of choices, adherence to choices, and single-round admission. The score ranking is prioritized, and the order of preferences in the application is then considered. Here is an example to illustrate this process:

Student 1 (S1) scored 626 points, and Student 2 (S2) from the same PA scored 625 points. S1's preference college order in the application is: X, Y, Z; S2's order is: Y, X, Z. The application submission and college admission process proceeds as follows:

\begin{itemize}

\item The machine reviews S1's application before S2 due to their higher score. A1's three preferred colleges (X, Y, Z) was reviewed. S1's first preference is X, but if X is already full for the approved college enrollment plan, the machine will move to the second preference, Y. Assuming Y still has one last slot, S1 will be admitted to Y according to their second preference.

\item After processing S1's application, the machine will review S2's choices in order. S2's first preference is Y, but since the last slot at Y has been filled by S1, who had a higher score, the machine will move on to the second preference, X. Since X is also full, the machine will then move to their third preference, Z. If Z is not full, S2 will be admitted to Z. If Z is also full, S2 will not be admitted by any university in this phase. 

\end{itemize}

One student can only be admitted to one institution, in comparison to colleges outside of mainland China where an student can be admitted to multiple colleges. This policy has been so for dozens of years due to the huge number of students and colleges. If a student is not satisfied with their admission offer and chooses not to accept it, they may return to high school, retake GaoKao exams in the following year, and reapply through the same process again.




\subsection{Demographics Information of Family Groups}

Table \ref{demo_info} and Table \ref{demo_info_expert} shows the demographics information for our participants.

\begin{sidewaystable}
    \vspace{16cm}
    \centering
        \caption{Demographic Information of Family Groups. PF1-PF7 represents family members in family-children pairs, PK1-PK7 represents children in family-children pairs. 
    K1-K6 represents individual children, F11-F14 represents individual family members.}
    \resizebox{\linewidth}{!}{
    \begin{tabular}{llp{1.cm}llllp{2.cm}lp{1.5cm}lp{1.5cm}p{1.5cm}}
    \toprule
\textbf{PID} & \textbf{Role}  & \textbf{Age Group} & \textbf{PA Region} & \textbf{Occupation} & \textbf{Education Level} &\textbf{PID} & \textbf{High School Track} & \textbf{Gender} & \textbf{College Tier} & \textbf{Major}         & \textbf{Year of Admission} & \textbf{Children 1st-Gen. College?} \\
\midrule
\textit{Family-children pairs} &                             &                                 &          &                               &                                           &                               &                                               &                                    &                    &                             &                                &                       
\\ \midrule
PF1                              & Father                            & 41-45                                  & Beijing         & Researcher                              & PhD                                          & PK1                              & Humanities                                              & F                                   & Others                   & Architectural Design                           & 2024                               & No                         \\
PF2                              & Mother                           & 51-55                                  & Hebei           & Cashiers                                & Junior High School                                           & PK2                              & Sciences                                              & M                                   & 普通一本                     & Mechanical Engineering                             & 2024                               & Yes                          \\
PF3                              & Uncle                             & 46-50                                  & Shaanxi         & Consultant                              & Graduate Student                                          & PK3                              & Sciences                                              & M                                   & 普通一本                     & Urban Planning                           & 2024                               & Yes                          \\
PF4                              & Mother                           & 41-45                                  & Ningxia         & Engineer                                    & Undergraduate                                           & PK4                              & Humanities                                              & F                                   & 二本                       & Chinese                        & 2023                               & No                         \\
PF5                              & Mother                           & 41-45                                  & Beijing         & Finance                                      & Graduate Student                                          & PK5                              & Humanities                                              & M                                   & 二本                       & Economics                           & 2023                               & No                         \\
PF6                              & Mother                           & 46-50                                  & Anhui           & Factory Worker                                   & Technical Secondary School                                           & PK6                              & Sciences                                              & F                                   & 985  & Engineering Mechanics       & 2023           & Yes                          \\
PF7                              & Mother                           & 46-50                                  & Beijing         & Architect                                      & Undergraduate                                           & PK7                              & Sciences                                              & M                                   & 211                      & Economics                           & 2024                               & No                         \\ \midrule
\textit{Individual children} &                             &                                 &          &                               &                                           &                               &                                               &                                    &                    &                             &                                &                       
\\ \midrule
                                 &                                   &                                        & Henan           & Farmer                                      & Not disclosed                                           & K1                               & Humanities                                              & M                                   & Others                       & Design                         & 2024                               & Yes                          \\
                                 &                                   & 46-50                                  & Innermongolia   & Self-employed                                      & Technical Secondary School                                           & K2                               & Sciences                                              & M                                   & 985                      & Computer Science                           & 2024                               & Yes                          \\
                                 &                                   & 46-50                                  & Hebei           & Middle school teacher                   &                                              & K3                               & Sciences                                              & F                                   & 985                      & Chemistry                             & 2021                               & No                         \\
                                 &                                  & 46-50                                  & Henan           & University instructor                   & Graduate Student                                         & K4                               & Sciences                                              & M                                   & 985                      & Economics                           & 2023                               & No                         \\
                                 &                                   &                                        & Innermongolia   &                                         &                                              & K5                               & Sciences                                              & M                                   & 985                      & Physics                        & 2022                               & Yes                       \\
                                 &                                   &                                        & Innermongolia   & Doctor                                         & Undergrad                                            & K6                               & Sciences                                              & M                                   & 985                      & Chemical Engineering Process                         & 2024                               & No                       \\ \midrule
\textit{Individual family members} &                             &                                 &          &                               &                                           &                               &                                               &                                    &                    &                             &                                &                       
\\ \midrule
                                 
F11                              & Mother                            & 41-45                                  & Henan           & Government Job                                     & Junior College                                           &                                 & Sciences                                              &                                     & 985                      & Undecided           & 2024                               & Yes                          \\
F12                              & Mother                           & 56-60                                  & Henan           & Administration                                      & Junior College                                           &                                  & Sciences                                              &                                     & 985                      & Life Sciences (Honors Program)                     & 2024                               & Yes                          \\
F13                              & Father                            & 46-50                                  & Henan           & Government Job                                    & Junior College                                           &                                  & Sciences                                              &                                     & 985                      & Computer Science                            & 2024                               & Yes                          \\
F14                              & Father                            & 46-50                                  & Henan           & Manufacturing Industry                                   & Undergraduate                                           &                                  & Sciences                                              &                                     & 985                      & Agricultural Engineering & 2024       & No                        \\
\bottomrule
\end{tabular}}%
    \label{demo_info}
\end{sidewaystable}

\newpage
\begin{table}[]
    \centering
        \caption{Demographic Information of Experts}
    \begin{tabular}{llllll}
\toprule
\textbf{PID} & \textbf{Gender} & \textbf{Age} & \textbf{Major Occupation}         & \textbf{PA Region} & \textbf{Number of Students Helped} \\
\midrule
E1           & M               & 36-40        & Self-Employed Admission Consultor & Shandong        & 3-10                                            \\
E2           & M               & 51-55        & University Professor and Employee & Beijing         & 10+                                             \\
E3           & M               & 46-50        & Middle School Teacher             & Chongqing       & 10                                              \\
E4           & M               & 56-60        & University Professor and Employee & Jiangsu         & 30yrs, 50+                                      \\
E5           & M               & 31-35        & Afterschool Program Teacher       & Tianjin         & 8yrs, 150                                       \\
E6           & F               & 31-35        & University Professor and Employee & Beijing         & 1-3                                             \\
E7           & M               & 31-35        & University Professor and Employee & Hebei           & 15                                              \\
E8           & M               & 41-45        & Software Engineer                 & Zhejiang        & 10 \\
\bottomrule                                            

\end{tabular}%

    \label{demo_info_expert}
\end{table}

\begin{figure}
    \centering
    \begin{subfigure}{0.21\textwidth}
        \centering
        \includegraphics[width=\textwidth]{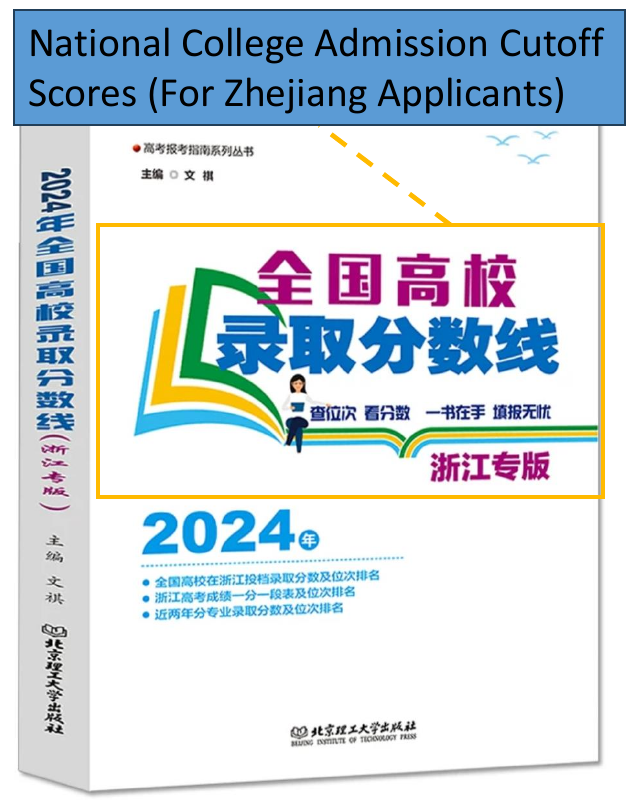}
        
        \Description{Front matter of a book}
        \label{fig:cutoffbook:title}
    \end{subfigure}
    \hfill
    \begin{subfigure}{0.38\textwidth}
        \centering
        \includegraphics[width=\textwidth]{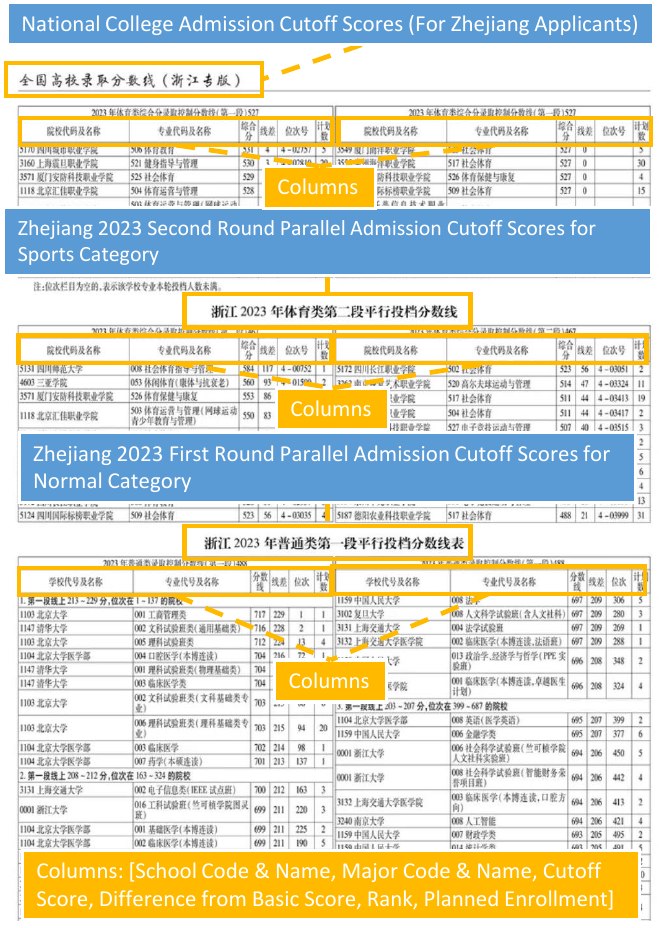}
        
        \Description{A book page of cutoff scores per school per major}
        \label{fig:cutoffbook:page1}
    \end{subfigure}
    \hfill
    \begin{subfigure}{0.39\textwidth}
        \centering
        \includegraphics[width=\textwidth]{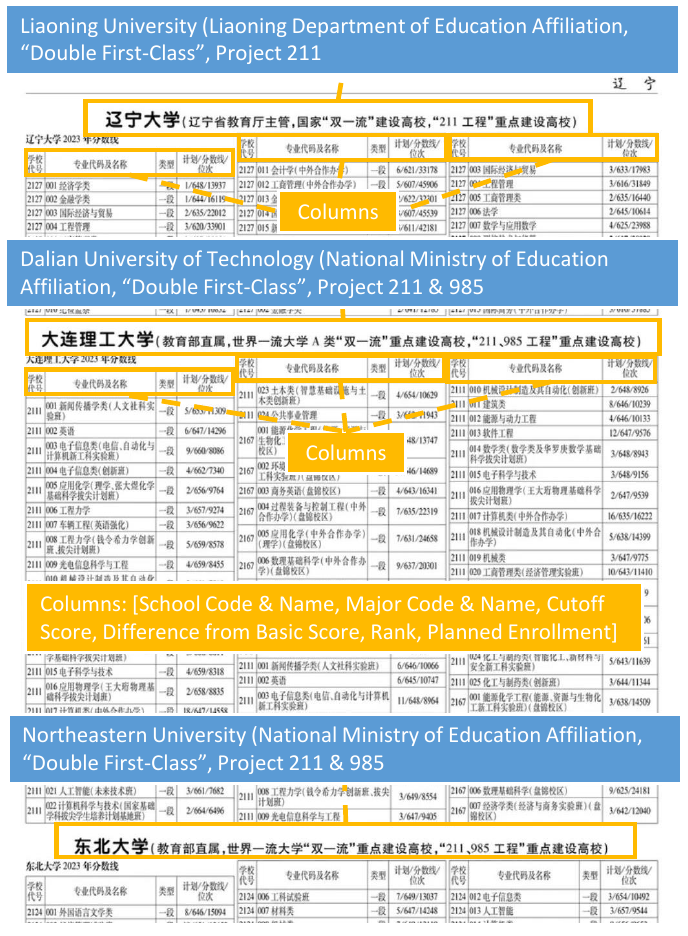}
        \Description{A book page of cutoff scores per school per major}
        \label{fig:cutoffbook:page2}
    \end{subfigure}

    \caption{Example of official guidebook for GaoKao college applications (left), and example contents in the book (middle and right). The book is over 200 pages with similar contents shown above.}
    
    \label{fig:cutoffbook}
\end{figure}





    
